\documentclass[twocolumn,pre,aps,showpacs,amsmath,amssymb,superscriptaddress]{revtex4-1}

\usepackage{hyperref}
\usepackage{amsmath,amssymb}
\usepackage{amsfonts,amsthm}
\usepackage{graphics}
\usepackage{graphicx}
\usepackage{dcolumn}
\usepackage{color}
\usepackage{bm}

\usepackage[normalem]{ulem}

\begin{document}

\title{The role of inhibitory neuronal variability in modulating phase diversity between coupled networks}

\author{Katiele V. P. Brito}
\affiliation{Instituto de F\'{\i}sica, Universidade Federal de Alagoas, Macei\'{o}, Alagoas 57072-970, Brazil}
\affiliation{Instituto de F\'{\i}sica Interdisciplinar y Sistemas Complejos IFISC (CSIC-UIB), Universitat de les Illes Balears, Campus UIB, E-07122 Palma de Mallorca, Spain}

\author{Joana M. G. L. Silva}
\affiliation{Instituto de F\'{\i}sica, Universidade Federal de Alagoas, Macei\'{o}, Alagoas 57072-970, Brazil}

\author{Claudio R. Mirasso}
\affiliation{Instituto de F\'{\i}sica Interdisciplinar y Sistemas Complejos IFISC (CSIC-UIB), Universitat de les Illes Balears, Campus UIB, E-07122 Palma de Mallorca, Spain}

\author{Fernanda S. Matias}
\thanks{fernanda@fis.ufal.br}
\affiliation{Instituto de F\'{\i}sica, Universidade Federal de Alagoas, Macei\'{o}, Alagoas 57072-970, Brazil}

\begin{abstract}

Neuronal heterogeneity, characterized by the presence of a multitude of spiking neuronal patterns, is a widespread phenomenon throughout the nervous system. In particular, the brain exhibits strong variability among inhibitory neurons. 
Despite the huge neuronal heterogeneity across brain regions, which in principle could decrease synchronization, cortical areas coherently oscillate during various cognitive tasks.
Therefore, the functional significance of neuronal heterogeneity remains a subject of active investigation. 
Previous studies typically focus on the role of heterogeneity in the dynamic properties of only one population.
Here, we explore how different types of inhibitory neurons can contribute to the diversity of the phase relations between two cortical areas. 
This research sheds light on the potential impact of local properties, such as neuronal variability, on communication between distant brain regions.
We show that both homogeneous and heterogeneous inhibitory networks can exhibit phase diversity and nonintuitive regimes such as anticipated synchronization (AS) and phase bistability.
It has been proposed that the bi-stable phase could be related to bi-stable perception, such as in the Necker cube.
Moreover, we show that heterogeneity enlarges the region of zero-lag synchronization and bistability.
We also show that the parameter controlling inhibitory heterogeneity modulates the transition from the usual delayed synchronization regime (DS) to AS. 
Finally, we show that the inhibitory heterogeneity drives the internal dynamics of the free-running population. Therefore, we suggest a possible mechanism to explain when the DS-AS transition occurs via zero-lag synchronization or bi-stability.
\end{abstract}
\maketitle

\section{Introduction}

The binding by synchrony hypothesis~\cite{Singer99} 
suggests that our perception is related to the synchronization of the activity of groups of neurons responsible for different image attributes (color, shape, speed). The fact that groups of distant neurons fire synchronously, despite their significant heterogeneity, enables the integration of various image attributes, allowing us to recognize an object as a unified whole: This is an apple!
More recently, alternative theories have proposed that neural oscillations (as described by the \textbf{communication through coherence} hypothesis~\cite{Fries05,Bastos15}) and \textbf{phase diversity}~\cite{Maris13,Maris16} play crucial roles in neuronal communication processes, further expanding our understanding of how the brain coordinates complex information.

 Inhibitory neurons are essential to cognitive functions and display remarkable variability.~\cite{di2021optimal}
 Diversity is a distinctive feature of the nervous system neurons as observed in various experiments~\cite{Angelo12, Beaulieu93, Markram04}.
 However, computational models, which often reduce neuronal networks to homogeneous  units~\cite{borges2020,nawrot2008measurement,rossi2020,burkitt2006review,borges2020influence}, fail to fully capture the implications of neuronal diversity on cortical processes~\cite{Angelo12,Tripathy13,Beaulieu93,Markram04}.
Many studies have investigated the role of neuronal heterogeneity in  various neuronal processes, including synchronization~\cite{rossi2020,di2021optimal},
coding capabilities~\cite{Golomb93, Shamir06, Perez10, Mejias12, Mejias14}, self-sustained activity~\cite{borges2020}, and persistent activity related to cognitive functions like working memory~\cite{Renart03, Hass19,model2003cellular}.
However, the functional implications of neuronal heterogeneity remain an active area of ongoing research~\cite{Marsat10,Padmanabhan10,Savard11,Angelo12,Tripathy13}. 

Previous studies typically focussed on the role of heterogeneity in the dynamical properties of a single population~\cite{rossi2020,di2021optimal,Borges2020}.
Here, we are interested in understanding how inhibitory neuronal variability influences the phase relation between two unidirectionally coupled networks.
The typical synchronized regime that emerges from the interaction of two unidirectionally coupled systems exhibits a positive phase lag in which the sender also acts as the leader. This regime is usually called delayed synchronization (DS) or lagged synchronization.
However, it was shown that the sender-receiver configuration can also synchronize with a negative phase lag if the system is governed by specific equations where the receiver is subject to delayed self-feedback.~\cite{Voss00}:
\begin{eqnarray}
\label{eq:voss}
\dot{\bf {S}} & = & {\bf f}({\bf S}(t)), \\
\dot{\bf {R}} & = & {\bf f}({\bf R}(t)) + {\bf K}[{\bf S}(t)-{\bf R}(t-t_d)]. \nonumber 
\end{eqnarray}
The stable solution ${\bf R}(t)={\bf S}(t+t_d)$ characterizes the counter-intuitive situation in which the receiver leads the sender, and it is called anticipated synchronization (AS). This means that the activity of the receiver system predicts that of the sender by an amount of time $t_d$.
In the last 20 years, this solution has been extensively studied in physical systems both theoretically~\cite{Voss00,Voss01b,Voss01a,Ciszak03,Masoller01,HernandezGarcia02,Sausedo14} and
experimentally~\cite{Sivaprakasam01,Ciszak09,Tang03}.

It has been shown that AS can also occur if the delayed self-feedback is replaced by different parameter mismatches at the receiver~\cite{Kostur05,Pyragiene13,Simonov14}, a faster internal dynamics of the receiver~\cite{Hayashi16,Dima18,Pinto19,DallaPorta19}, as well as by an
inhibitory loop mediated by chemical synapses~\cite{Matias11,Matias16,Matias17,Mirasso17}.
 Moreover, when the feedback is not hard-wired in the equation but emerges from system dynamics, two neuronal populations can exhibit phase diversity and a transition from AS to DS through zero-lag synchronization, induced by synaptic properties~\cite{Matias14,DallaPorta19}. Furthermore, AS has been observed in electrophysiological data of nonhuman primates ~\cite{Matias14} and human EEG~\cite{carlos2020anticipated} while performing a cognitive task. In experiments, unidirectional coupling can be assessed using Granger causality and may be associated with either a positive or negative phase difference between brain areas~\cite{Matias14,Brovelli04,Salazar12,carlos2020anticipated}.

 More recently, it has been shown that a sender-receiver network model can also exhibit phase bistability between DS and AS regimes~\cite{machado2020phase,brito2021neuronal}. It has been suggested that the bi-stable phase may be linked to bi-stable perception, as exemplified by the Necker cube. 
 An experimental study using magnetoencephalography has demonstrated a strong phase relationship between brain signals during auditory bistable perception in humans~\cite{Kosem16}. 
 In this study, volunteers listened to bistable speech streams that could be perceived as two distinct word sequences, such as "fly-fly-fly" or "life-life-life." The authors reported that the phase difference consistently reflected the word sequence perceived by the participants.

Therefore, here, we investigate the role of inhibitory neuronal variability in a two-network model that can exhibit these two nonintuitive regimes: anticipated synchronization and phase bistability between AS and DS.
We are not aware of other studies relating the heterogeneity of inhibitory neurons and the phase relations between two coupled populations, but only the heterogeneity of excitatory neurons~\cite{brito2021neuronal}.
In Sec.~\ref{model}, we describe the neuronal population model as well as the parameters that we use to manipulate neuronal heterogeneity.
In Sec.~\ref{results}, we report our results, showing that the motif can exhibit phase-locking regimes with positive, negative, and zero phases, as well as a bistable regime that alternates from AS to DS. We also show that the internal dynamics of the receiver influence the type of transition from DS to AS via zero lag or bistability. 
Concluding remarks and a brief discussion on the significance of our findings for neuroscience are presented in Sec.~\ref{conclusions}.

\begin{figure}[t]%
\includegraphics[width=0.8\columnwidth,clip]{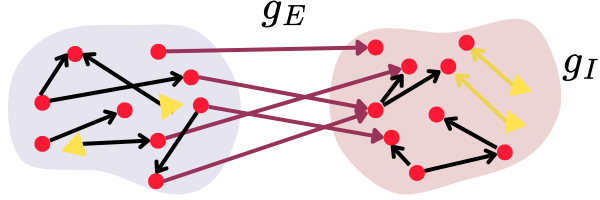}
\caption{\label{fig1:SenderReceiver} Schematic representation of two unidirectionally coupled neuronal populations, displaying a sender-receiver motif. Each population is composed of hundreds of neurons, with the proportions of different excitatory and inhibitory neuron types at the receiver (R) population varying throughout the study. The inhibitory feedback at R is regulated by the synaptic conductance $g_I$, while the coupling between the sender and receiver is mediated by excitatory synaptic conductance $g_E$. These conductances, along with the parameters controlling excitatory and inhibitory neuronal heterogeneity ($X$ and $X_i$), modulate the dynamics of the motif.}
\end{figure}%

\section{The model}
\label{model}

\begin{figure}[!t]%
\includegraphics[width=1.0\columnwidth,clip]{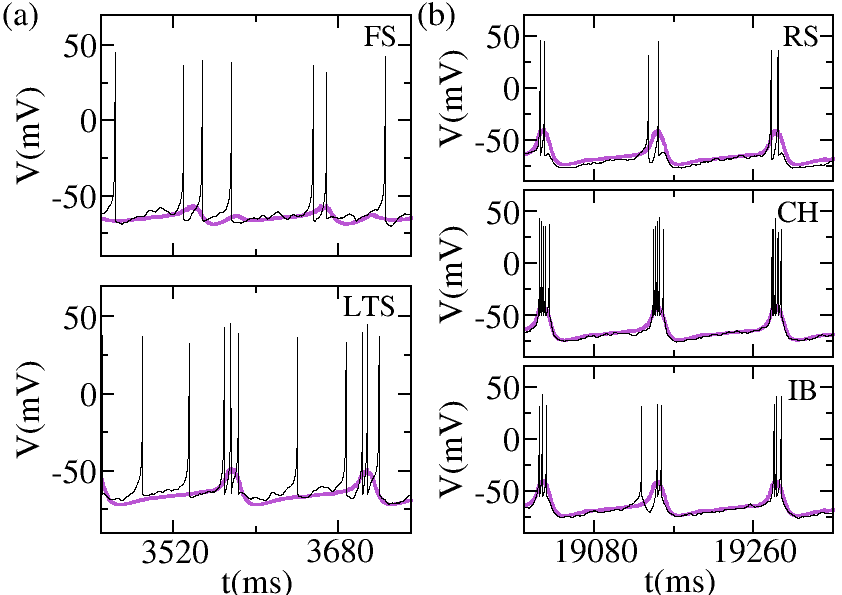}
\caption{\label{fig:neurontypes}
Examples of neuronal time series illustrate the heterogeneity of individual neurons' firing patterns at the receiver population during a phase-locking regime.
(a) Two inhibitory neurons: a fast-spiking neuron (FS) and a low-threshold-spiking neuron (LTS). (b) Three excitatory neurons: a regular-spiking neuron (RS), a chattering neuron (CH), and an intrinsically bursting neuron (IB). 
}
\end{figure}%

\begin{figure}[!t]%
\includegraphics[width=1.0\columnwidth,clip]{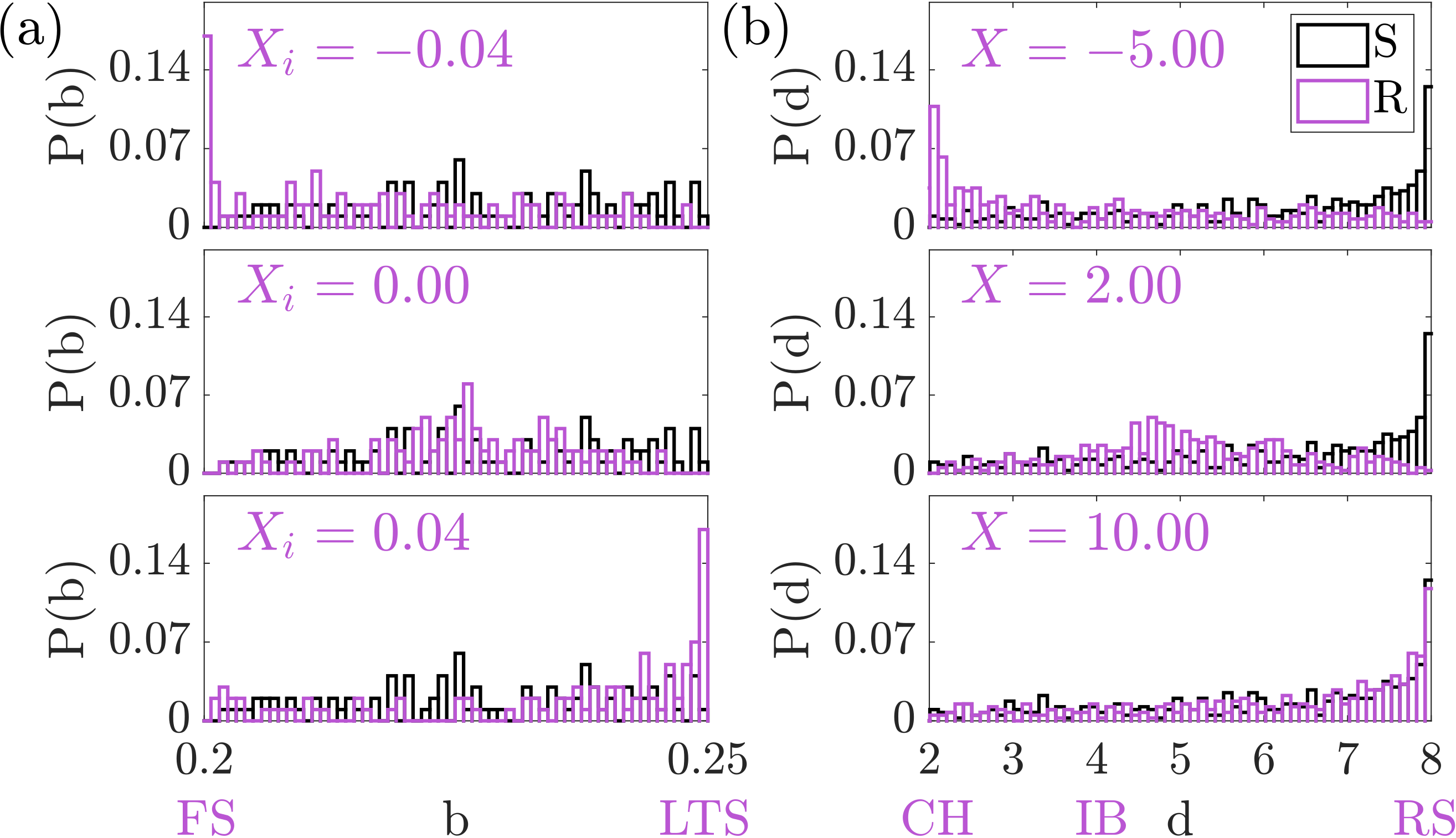}
\caption{\label{fig:neuronprobability}
Characterizing the parameters controlling neuronal heterogeneity:$X$ and $X_i$.
The plots show the probability distribution of different types of (a) inhibitory and (b) excitatory neurons in each population. 
$X_i$ is the parameter controlling inhibitory heterogeneity and determining the distribution of the values of $b$ in Eqs. (\ref{eq:a}), (\ref{eq:b}).
(a) For $X_i = -0.04$ the distribution is predominantly composed of fast-spiking neurons (FS, $a = 0.10$ and $b = 0.20$) while for $X_i = 0.00$, the parameter $b$ is at a mid-range value of both FS and low-threshold spiking (LTS) types. At $X_i = 0.04$, the distribution shifts predominantly to LTS neurons ($a = 0.02$ and $b = 0.25$). (b) For excitatory neurons, if $X = -5.00$, the distribution favors $d = 2$ and $c=-50$, indicating a prevalence of chattering (CH) neurons. At $X = 2.00$, $d$ peaks around 5, with more intrinsically bursting (IB) neurons ($d = 4, c = -55$), and at $X = 10.00$, regular spiking (RS) neurons ($d = 8, c = -65$) dominate. 
}
\end{figure}%

\subsection{Fundamental components of the networks: nodes as spiking neuron models}

Our neuronal motif is modeled with two unidirectionally coupled cortical-like neuronal populations: a sender (S) and a receiver (R), see Fig.\ref{fig1:SenderReceiver}. Anatomical estimates for the neocortex suggest a standard ratio of $80 \%$ excitatory neurons and $20 \%$ inhibitory neurons. Therefore, we consider each population comprised of $400$ excitatory and $100$ inhibitory neurons. Each neuron is described by the Izhikevich model~\cite{Izhikevich00}:
\begin{eqnarray}
  \frac{dv}{dt} &=& 0.04v^2+5v+140-u +\sum_x I_{x}, \label{dv/dt}\\
  \frac{du}{dt} &=& a(bv-u), \label{du/dt}
\end{eqnarray}
where $v$ is the membrane potential, and $u$ is the recovery variable, accounting for activation of K$^+$ and inactivation of Na$^+$ ionic currents.
$I_x$ are the synaptic currents provided by the interaction with other neurons and external inputs. 
If $v\geq30$~mV, $v$ is reset to $c$ and $u$ to $u+d$. 
The mean membrane potential of each population is calculated as the average of all $N$ neuronal membrane potentials in that population: $V_x=\sum_{i} v_{x}^{i}/N$ ($x=$S, R).

The dimensionless parameters $a, b, c$ and $d$ determine the neuronal type.
This means that for an external constant current, the neuron's spiking properties are well known~\cite{Izhikevich00}.
For instance, inhibitory neurons can exhibit fast-spiking (FS) dynamics with $a = 0.10$ and $b = 0.20$ or low-threshold spiking (LTS) with $a = 0.02$ and $b = 0.25$. 
Both types share common parameters $c=-65$ and $d=2$. 
Figures~\ref{fig:neurontypes}(a) illustrate the time traces of an FS neuron in the upper panel and an LTS neuron in the bottom panel, both from the receiver population. 
Unlike the case of a constant external current, these neurons receive multiple synaptic inputs from other neurons, contributing to their dynamic behavior.
The dynamics of excitatory neurons can vary, exhibiting regular spiking (RS) behavior for $c = -65$ and $d = 8$, intrinsically bursting (IB) for $c = -55$ and $d = 4$, or chattering (CH) spiking for $c = -50$ and $d = 2$. 
All excitatory neurons have fixed $(a,b)=(0.02,0.2)$. Figures~\ref{fig:neurontypes}(b) illustrate the time traces of one excitatory neuron of each type: RS, CH, and IB. 

To model all the neurons in each population, the dimensionless parameters are typically randomly sampled as follows: $(a,b)=(0.02,0.2)$ and $(c,d)=(-65,8)+(15,-6)\sigma^2$ for excitatory neurons, and $(a,b)=(0.02,0.25)+(0.08,-0.05)\sigma$ and $(c,d)=(-65,2)$ for inhibitory neurons. 
$\sigma$ is a random variable drawn from a uniform distribution on the interval $[0,1]$~\cite{Izhikevich03,Izhikevich04a}. 
These values account for the natural variability of neuronal dynamics in cortical networks. 
Previous works about anticipated synchronization have followed this approach, maintaining these parameter values unchanged~\cite{Matias14,Matias15}. 
We use this distribution in our sender population, where the various neuron types are represented in black for the sender in Fig.~\ref{fig:neuronprobability}. 
In the case of excitatory neurons, this distribution favors a higher proportion of regular spiking neurons, while for inhibitory neurons, it results in a more balanced distribution between FS and LTS types.

\subsection{Controlling the neuronal heterogeneity}
Recently, we have investigated the effect of changing the distribution of the parameters $c$ and $d$ only for the excitatory neurons~\cite{brito2021neuronal}. 
To investigate the impact of neuronal variability in the receiver population, we use different values of $(c,d)$ for the excitatory neurons in R as in Ref.~\cite{brito2021neuronal}:

\begin{eqnarray}
  c &=& -55-X+[(5+X)\sigma^2]-[(10-X)\sigma^2], \label{eq:c} \\
  d &=& 4  +Y-[(2+Y)\sigma^2]+[(4-Y)\sigma^2]. \label{eq:d}
\end{eqnarray}
For excitatory neurons, we ensure a gradual transition in the distribution of neuronal types, from regular spiking to chattering, passing through intrinsically bursting neurons, with the relationship $Y = \frac{2X}{5}$."
As we vary parameter $X$ from -5 to 10, the composition of neuronal types within the R population changes, as illustrated in 
Fig.~\ref{fig:neuronprobability}(b). 
For $X=-5$ the majority of the neurons are CH, whereas for $X=10$ most of the neurons are RS. 
This approach has been used before to understand the role of excitatory heterogeneity in AS~\cite{brito2021neuronal}.

To investigate the role of inhibitory heterogeneity within the receiver population, we consider distinct values for parameters $a$ and $b$ for the inhibitory neurons in the receiver population:

\begin{eqnarray}
a &=& 0.06 - X_i + [(0.04 + X_i)\sigma^2)] - [(0.04 - X_i)\sigma^2], \label{eq:a}\\
b &=& -0.625 a + 0.262. \label{eq:b}
\end{eqnarray}

These values determine the distribution of neuronal types from low-threshold spiking to fast-spiking neurons.
As we vary $X_i$ within the range of $-0.045$ to $0.045$, the peak probability $P(b)$ of finding a neuron with a specific $b$ value shifts from $b = 0.2$ (FS neurons) to $b = 0.25$ (LTS neurons). 
This indicates that the quantity of each neuron type within the R population varies as depicted in Fig.~\ref{fig:neuronprobability}. 
For $X_i=-0.04$ the majority of the neurons are FS, while for $X_i=0.04$ most neurons are LTS. 
Throughout the paper, we vary $X_i$ and $X$ to demonstrate how heterogeneity influences the phase relationships between the two populations. 

We also examine the effects of homogeneous networks on the diversity of phase relations. 
In these simulations, we fix all parameters $a$ and $b$ to have only fast-spiking inhibitory neurons (\textit{only-FS network}). or only low-threshold spiking inhibitory neurons (\textit{only-FS network}).
These two cases represent the extreme situations in which the histograms in Fig.~\ref{fig:neuronprobability} display all values as zero except one. 
The only-LTS case would be similar to $X_i=0.04$ but with $P(b=0.25)=1$.
Similarly, the only-RS case would be close to the $X=10$ case but with $P(d=8)=1$.

\begin{figure}[!ht]%
\includegraphics[width=1.0\columnwidth,clip]{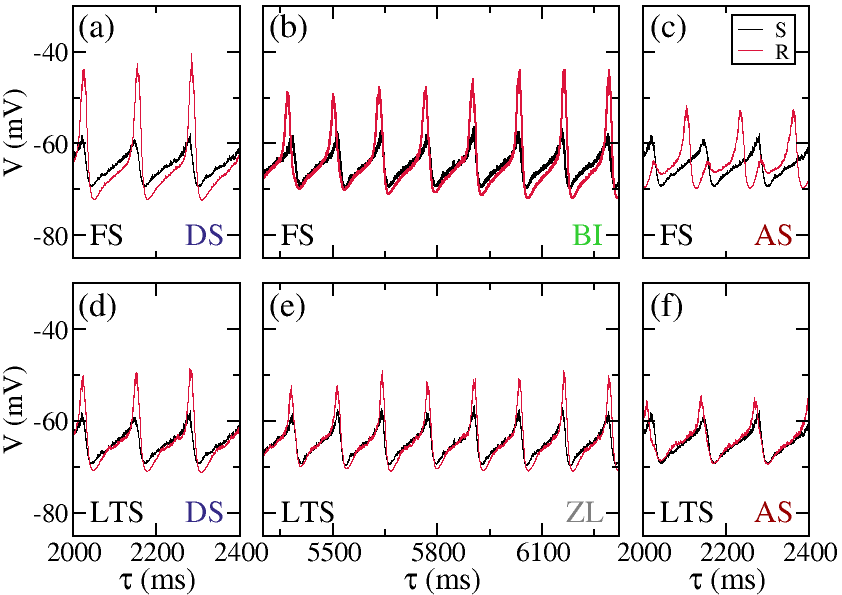}
\caption{\label{fig:Vmemb} 
Revealing phase relations diversity in our motifs. Time traces illustrate the membrane potentials $V_X$ of the sender (S) and receiver (R) populations under phase-locking and bistable regimes. 
(a)-(c) The top panels depict homogeneous inhibitory networks composed of only fast-spiking neurons (only-FS networks), which can exhibit delayed synchronization (DS) at $X$ = -5.00, anticipated synchronization (AS) at $X$ = -3.00, and a bistable regime concerning phase differences (BI) at $X$ = -4.00.  (d)-(f) The only-LTS network (bottom panel) can exhibit DS at $X$ = -1.00, AS at $X = 1.00$, and a zero-lag regime (ZL) at $X$ = 0.00, where the two populations fire synchronously without any temporal delay. The synaptic weights are $g_I = 5.00$~nS and $g_E = 0.50$~nS. 
}
\end{figure}%

\subsection{Network connectivity: each link is a chemical synapse}

\begin{figure}[!t]%
\begin{minipage}{8cm}
\includegraphics[width=1.00\columnwidth,clip]{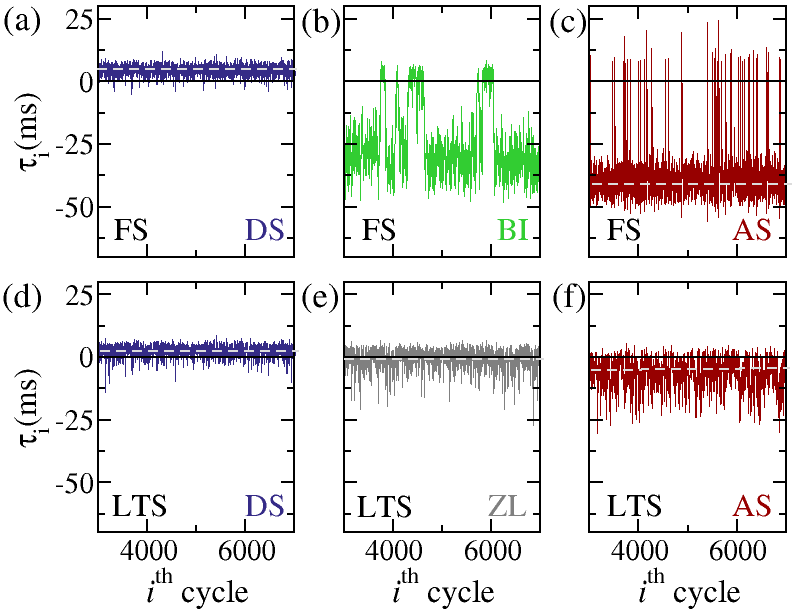}
\end{minipage}
\begin{minipage}{8cm}
\includegraphics[width=1.0\columnwidth,clip]{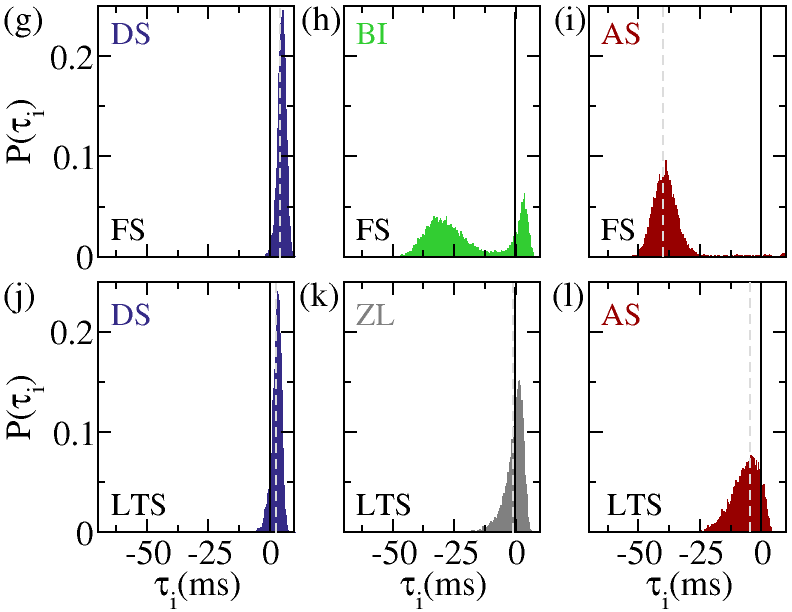}
\end{minipage}
\caption
{\label{fig:tau_cycle}
Illustrating the behavior of the time delays in different regimes. The time delay in each cycle $\tau_i$ \textit{versus} the $i-th$ cycle: (a)-(c) for the only-FS network, (d)-(f) for the only-LTS network.
(g)-(l) Histograms showing the probability distribution of the time delays $\tau_i$. Parameters are equivalent to the ones in the upper panel (a)-(f).
The positive average of the delays $\tau_i$ cycles denotes the DS regimes (blue), while the negative values of $\tau_i$ on average correspond to the AS regime (red). Phase bi-stability (green) is observed as an unpredictable alternation between DS and AS cycles. In the ZL condition (grey), the average of cycles is near zero, indicating synchronization with minimal phase difference. 
}
\end{figure}%

\begin{figure}[!t]%
\includegraphics[width=1.0\columnwidth,clip]{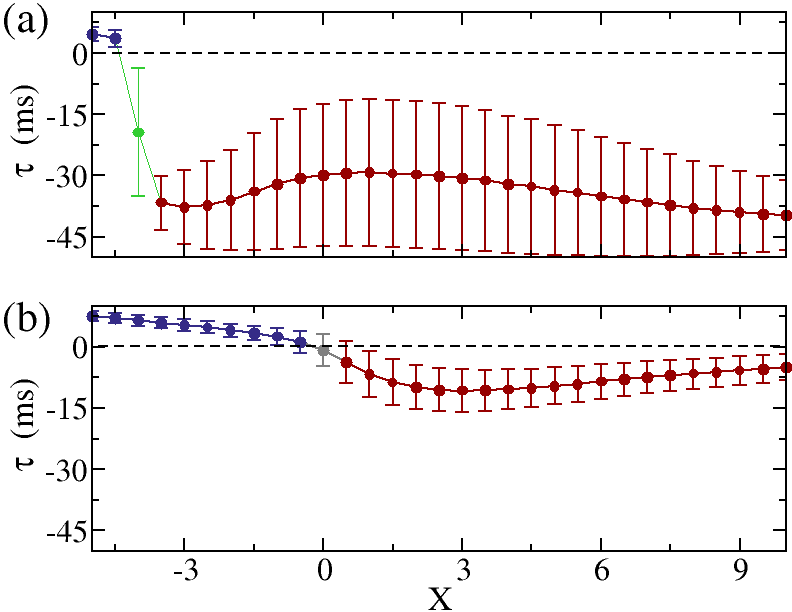}
\caption{\label{fig:FS_LTS_tau_X}
Comparison between the two different transitions from DS to AS: via bistability and via zero-lag synchronization. The mean time delay $\tau$ as a function of the parameter controlling neuronal heterogeneity $X$. (a) The only-FS network exhibits a bi-stable transition from DS to AS. (b) The only-LTS network presents a DS-AS transition via zero-lag. Error bars represent the standard deviation of $\tau$. The synaptic conductances are $g_i=5.00$~nS and $g_e=0.50$~nS.
}
\end{figure}%

The connections between neurons in each population are assumed to be
fast unidirectional excitatory and inhibitory chemical synapses
mediated by AMPA and GABA$_\text{A}$. The synaptic currents are given
by:
\begin{equation}
\label{Ix}
I_{x} = g_{x}r_{x}(v-V_x),
\end{equation}
where $x=E,I$ (excitatory and inhibitory mediated by AMPA and
GABA$_\text{A}$, respectively), $V_E=0$~mV, $V_I=-65$~mV, $g_{x}$ is the 
synaptic conductance maximal strengths  and $r_{x}$
is the fraction of bound synaptic receptors whose dynamics are given
by:
\begin{equation}
\label{drdt}
  \tau_x\frac{dr_{x}}{dt}=-r_{x} + D \sum_k \delta(t-t_k),\\
\end{equation}
where the summation over $k$ stands for pre-synaptic spikes at times
$t_k$. D is taken, without loss of generality, equal to $0.05$. The time decays are $\tau_{E}=5.26$~ms $\tau_{I}=5.6$~ms.  Each neuron is subject to an independent noisy spike train described by a Poisson distribution with rate $R$. The input mimics excitatory synapses (with conductances $g_{P}$) from $n$ pre-synaptic neurons external to the population,
each one spiking with a Poisson rate $R/n$ which, together with a constant external current $I_c$,  
determine the main frequency of mean membrane potential of each population. Unless otherwise stated, we have employed $R=2400$~Hz and $I_c=0$. 
The equations were integrated using the Euler method with a time step of $0.05$~ms.

Connectivity within each population randomly targets 10\% of the neurons, with excitatory conductances set at $g^{S}_{E} = g^{R}_{E} = 0.5$~nS. Inhibitory conductances are fixed at the sender population $g^{S}_{I} = 4.0$~nS and $g_I$ at the receiver population is varied throughout the study (see Fig.~\ref{fig1:SenderReceiver}).
Each neuron at the R population receives 20 fast synapses (with conductance $g_{E}$) from randomly selected excitatory neurons of the S population.

\section{Results}
\label{results}

\subsection{Investigating phase relations diversity}

We start by mimicking the oscillatory activity in cortical regions. We simulated the sender population in such a way that the
mean membrane potential $V_S$ oscillates with a mean period around $T_S\approx 130$~ms with neuronal heterogeneity in both excitatory and inhibitory neurons as described in Sec.~\ref{model}. Black curves in Fig.~\ref{fig:Vmemb} show $V_S$ as a function of time to illustrate the collective activity of all neurons in the sender populations for the same set of parameters.

In order to investigate the role of inhibitory neurons in the synchronization properties of connected populations, we start by investigating whether a receiver population with a homogeneous inhibitory network could synchronize with the sender. In what follows, we show the phase relation diversity between the S and R populations
when the inhibitory neurons of the receiver population are all fast-spiking neurons (\textit{only-FS network}).
Red curves in Fig.~\ref{fig:Vmemb}(a)-(c) show the time traces of $V_R$ for three different set of parameters for the only-FS case.

Since $V_x$ is noisy, we average within a sliding window of width $5-8$~ms to obtain a smoothed signal, from which we can extract 
the peak times $\{t^{x}_{i}\}$ (where $i$ indexes the i-th peak). 
The times of these peaks are essential indexes to characterize the phase relation between the two populations.
The period of a given population in each cycle is thus $T^{x}_i \equiv t^{x}_{i+1}-t^{x}_{i}$. 
For a sufficiently long time series, we compute the mean period
$T_x$ and its variance.  Similarly, we calculate the time delay between the oscillatory activity of the two populations in each cycle
$\tau_i=t^{R}_{i}-t^{S}_{i}$. Then, if $\tau_i$ obeys an unimodal distribution,  we calculate $\tau$ as the mean
value of $\tau_i$ and $\sigma_{\tau}$ as its variance.  
In such cases, the populations exhibit
oscillatory synchronization with a phase-locking regime. 
We indistinguishably use the term phase difference or time delay since it is always possible to associate both: $\phi_i=2\pi \tau_i/ T_{S}$. In all those calculations, we discard the
transient time.

\subsubsection*{The Delayed Synchronization (DS) regime}

 The only-FS case can present the usual delayed synchronization (DS) regime for $g_i=5.0$~nS and $X=-5.0$.
The mean membrane potential $V_x$ ($x=$S, R) of each population 
is shown in Fig.~\ref{fig:Vmemb}(a) for this illustrative example of DS.~\ref{fig:tau_cycle}
The receiver also oscillates with $T_S=T_R=130$~ms and presents peaks after the receiver as shown by the red curves in Fig.~\ref{fig:Vmemb}(a). We can note that $X=-5.0$ corresponds to a neuronal distribution in which the majority of excitatory neurons are chattering (CH, see Figures ~\ref{fig:neuronprobability} and ~\ref{fig:neurontypes}).

The phase-locking regimes can be characterized by the mean time delay $\tau$ and its standard deviation. If the mean time delay is positive ($\tau>0$), which indicates that the sender population leads the receiver, the system is in a DS regime.
The left panels of Fig.~\ref{fig:tau_cycle} show
examples of DS. For the only-FS case illustrated in Figs.~\ref{fig:Vmemb}(a) and ~\ref{fig:tau_cycle})(a),(g), the peak of the receiver population occurs on average $4.7$~ms  after the peak of the sender. This value is close to the magnitude of the synaptic timescales (as mentioned in
Sec.~\ref{model} $\tau_E = 5.26$~ms).
The distribution of $\tau_i$ in each cycle is unimodal for phase-locking regimes, as we can see in Figs.~\ref{fig:tau_cycle}(g),~(i)-(l).

\subsubsection*{The Anticipated Synchronization (AS) regime}

As we change the excitatory neuronal heterogeneity to have more intrinsically bursting and fast-spiking neurons, the system can exhibit a nonintuitive phase-locking regime called anticipated synchronization in which the receiver's peak occurs on average before the sender's peak.
For $X>-3.0$ in the only-FS network, the receiver also oscillates with $T_S=T_R=130$~ms but presents peaks before the receiver. See an example of the mean membrane potential of the S and R population in Fig.~\ref{fig:Vmemb}(c) for $X=-3.0$.
Fig.~\ref{fig:tau_cycle}(c)
shows $\tau_i$ in each cycle $i$, whereas Fig.~\ref{fig:tau_cycle}(i) exhibits
 the probability distribution of $tau_i$ for this illustrative example of AS in the only-FS network, which presents a negative mean time delay ($\tau= -37.77$~ms).

\subsubsection*{Phase-bistability between AS and DS}

A classic example of a bistable system is the Kramers problem, which describes how a system moves in a \textit{double-well potential} under the influence of noise. Imagine a ball sitting in one of two valleys (\textit{the two wells}). With enough noise or disturbances, the ball can move from one valley to another, representing a switch between the two states. The time the system (or the ball) spends in each valley before jumping to the other depends on how much noise is present. If the noise is low, it takes longer to switch between states, but as the noise increases, transitions between the two regions become more frequent. The probability distribution of the ball's position would be a bi-gaussian with peaks at the two valleys.

Here, we show that depending on the model parameters, the system can present a bistable regime (BI) between a DS and an AS regime characterized by a bi-gaussian distribution of the time-delays $\tau_i$. In particular, we show an example of phase bistability for the only-FS network with $X=-4$ in Fig.~\ref{fig:Vmemb}(b) and Figs. ~\ref{fig:tau_cycle}(b) and (h). 

The peaks of the mean membrane potential of the senders occur right before the peaks of the receiver for a few cycles. Therefore, the time delay $\tau_i$ is positive for these cycles, with well-defined positive mean values and standard deviation, similar to a DS regime for a certain number of cycles. Then, the system can suddenly switch to different dynamics in which the peaks of the mean membrane potential of the S population occur after the peaks of the R population for a few other cycles.
Therefore,
$\tau_i$ is negative during these cycles, similar to an AS regime for a certain amount of time. By analyzing the
system only for a few oscillation periods, one could wrongly characterize the system in purely DS or AS regime. But then,
randomly, the system can jump again to a different attractor. Therefore, the histogram of $\tau_i$ is a bi-Gaussian with one positive and one negative peak as shown in Fig.~\ref{fig:tau_cycle}(b) and (h). In this regime, which is not a phase-locking regime anymore, the system cannot be simply characterized by the mean time delay $\tau$ and its standard deviation. 

As we slowly change $X$ the proportion of excitatory neurons, the system undergoes a transition from DS to AS via bistability. Fig.~\ref{fig:FS_LTS_tau_X}(a)) evidences this transition by showing the mean time delay as a function of $X$ and depicting its error bars
A similar transition from DS to AS mediated by the type of excitatory neurons has been reported before when the inhibitory networks of the sender and receiver have a similar distribution of neuronal types~\cite{brito2021neuronal}, but not for only-FS neurons at the receiver. 
This is the first verification that a homogeneous inhibitory network can present anticipated synchronization and phase bistability.

\begin{figure}[!t]%
\includegraphics[width=1.0\columnwidth,clip]{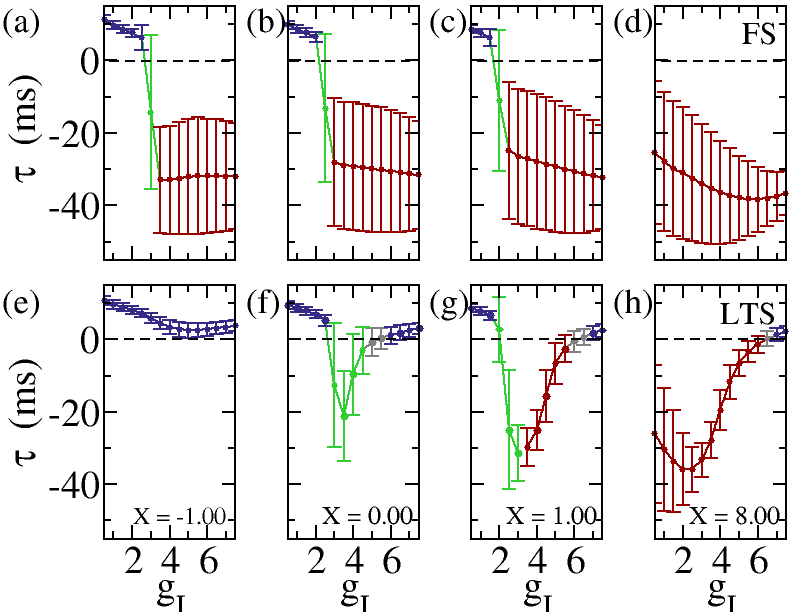}
\caption{\label{fig:FS_LTS_tau_gi}
Phase diversity is modulated by both inhibitory conductance and neuronal variability. The mean time delay as a function of inhibitory conductance $g_i$ for different values of $X$ (a)-(d) The only-FS network can present a transition from DS to AS via a bi-stable regime. (e)-(h) The only-LTS network can present different DS-AS transitions through bi-stable and ZL states.
}
\end{figure}%

\begin{figure}[!t]%
\includegraphics[width=1.0\columnwidth,clip]{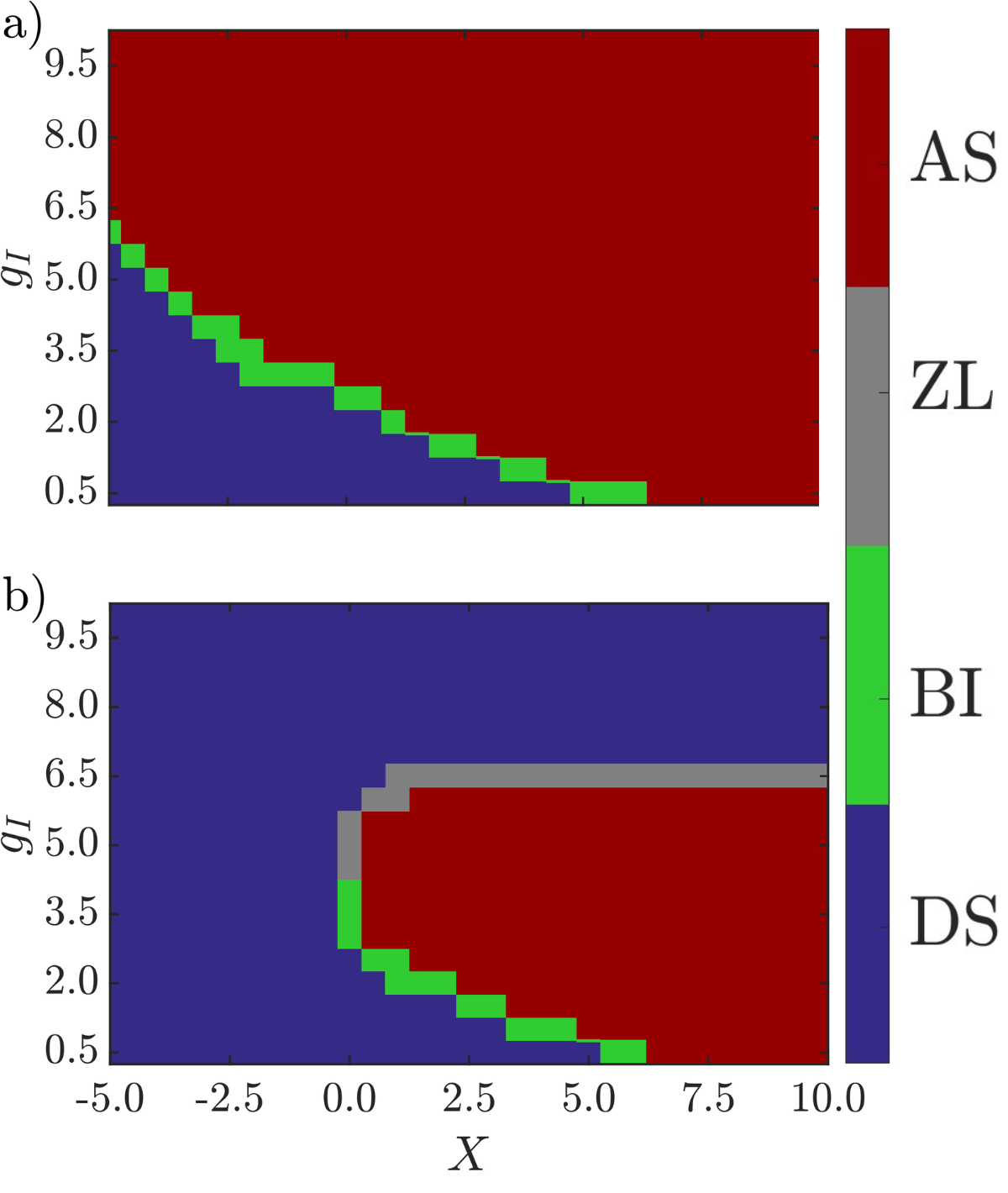}
\caption{\label{fig:FS_LTS_map_gi_X}
Scanning parameter space to visualize the phase relation
as a function of inhibitory synaptic conductance $g_I$ and the parameter controlling neuronal variability $X$ for (a) the only-FS network and (b) the only-LTS network.
The horizontal lines for $g_i=5.0$~nS are equivalent to the curves $\tau$ \textit{versus} $X$ in Fig. \ref{fig:FS_LTS_tau_X}.
Vertical lines for fixed values of $X$ ($X=-1.0$, $X=0.0$, $X=1.0$, and $X=8.0$) are equivalent to the curves $\tau$ \textit{versus} $g_i$ in Fig.\ref{fig:FS_LTS_tau_gi}.
}
\end{figure}%

\subsection{The role of inhibitory neurons: comparing only-FS and only-LTS homogeneous inhibitory networks}

In order to continue to explore the effects of inhibitory neurons in phase relations, we investigate the effects of inhibitory neuronal variability, comparing the previous results with the ones in another homogeneous inhibitory network. We compare our results of an only-FS inhibitory network with the case in which all inhibitory neurons of the receiver population are low-threshold spiking neurons: the \textit{only-LTS network}. 

For all the same parameters as before, the only-LTS network
also presents a DS regime for $X=-5$ and an AS regime for larger values of $X>0.25$.  
However, the transition from DS to AS not only occurs for different values of $X$ compared to the only-FS case, but it also occurs in a different way: via zero-lag synchronization (ZL) and not via bistability (BI).

In Fig.~\ref{fig:Vmemb}(d), (e), and (f), we can see the time series of the mean membrane potential for one example of each regime in the only-LTS network: DS ($X=-1.0$), ZL($X=0.0$), AS($X=1.0$). We also show the $\tau_i$ for cycle and its probability distribution in Fig.~\ref{fig:tau_cycle}(d)-(f) and (j)-(l).

Fig.~\ref{fig:FS_LTS_tau_X} show the mean time delay as a function of $X$, the parameter controlling the excitatory neuron type distribution, for different inhibitory networks: only-FS and only-LTS. These plots evidence the two different DS-AS transitions: via bistability and via zero-lag. 
It is worth mentioning that the mean time delay of anticipation in modulus can be larger for the only-FS network than for the only LTS. The discontinuity of $\tau$ along the AS-DS transition is related to the existence of the bistability. For the only-LTS network, the transition via ZL occurs smoothly with the peak of the histograms and the mean time delay changing continuously with $X$.


When we vary the connectivity properties of the inhibition, which is related to the feedback controlling AS, we can find even more phase diversity.
Fig.~\ref{fig:FS_LTS_tau_gi} shows the mean time delay as a function of the inhibitory conductance $g_I$ for different values of $X$.  The upper panels are for only-FS network and the lower panels are for the only-LTS case. 
In Fig.~\ref{fig:FS_LTS_tau_gi}(g), we can see that as we increase the inhibitory conductances, the system undergoes a DS to AS transition via bistability. However, if we increase $g_i$ even more, the system undergoes a new transition, returning to the delayed synchronization regime via zero-lag. 

Figs~\ref{fig:FS_LTS_map_gi_X}(a) and (b) show a colored map displaying a two-dimensional projection of the parameter space of our model ($X$,$g_I$) for only-FS and only-LTS networks.
 Each curve in Fig.~\ref{fig:FS_LTS_tau_X} corresponds to a horizontal line $g_i=5.0$~nS in Figs.~\ref{fig:FS_LTS_map_gi_X}(a) and (b).  
Each curve in Fig.~\ref{fig:FS_LTS_tau_gi} corresponds to a vertical line for $X=-1.0$, $X=0.0$, $X=1.0$, and $X=8.0$ in Figs.~\ref{fig:FS_LTS_map_gi_X}(a) and (b). 
The parameter space in which anticipated synchronization occurs is larger for only-FS neurons. The AS-DS transition occurs only via bistability. 
For only-LTS neurons, the region of DS is more extensive and the transitions can occur both via ZL or bistability.

Even for these homogeneous inhibitory networks, the system can present diverse phase relations. In particular, the inhibitory conductances can lead the system out of the phase-locking regime into a phase bistability.
Even though in other models, homogeneous inhibitory interneurons move the system toward highly synchronized dynamics~\cite{gast2024neural}, here, even homogeneous networks of inhibitory neurons can promote phase-bistability.
Furthermore, we show that populations with different types of inhibitory neurons show significant differences in the transition between phase-locking regimes.

\begin{figure}[!t]%
\begin{minipage}{8cm}
\includegraphics[width=0.98\columnwidth,clip]{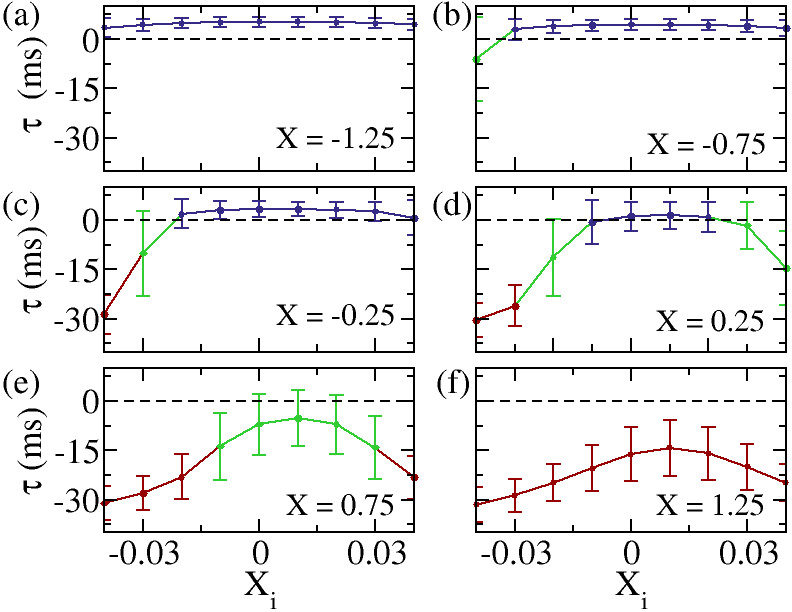}    
\end{minipage}
\begin{minipage}{8cm}
\includegraphics[width=0.98\columnwidth,clip]{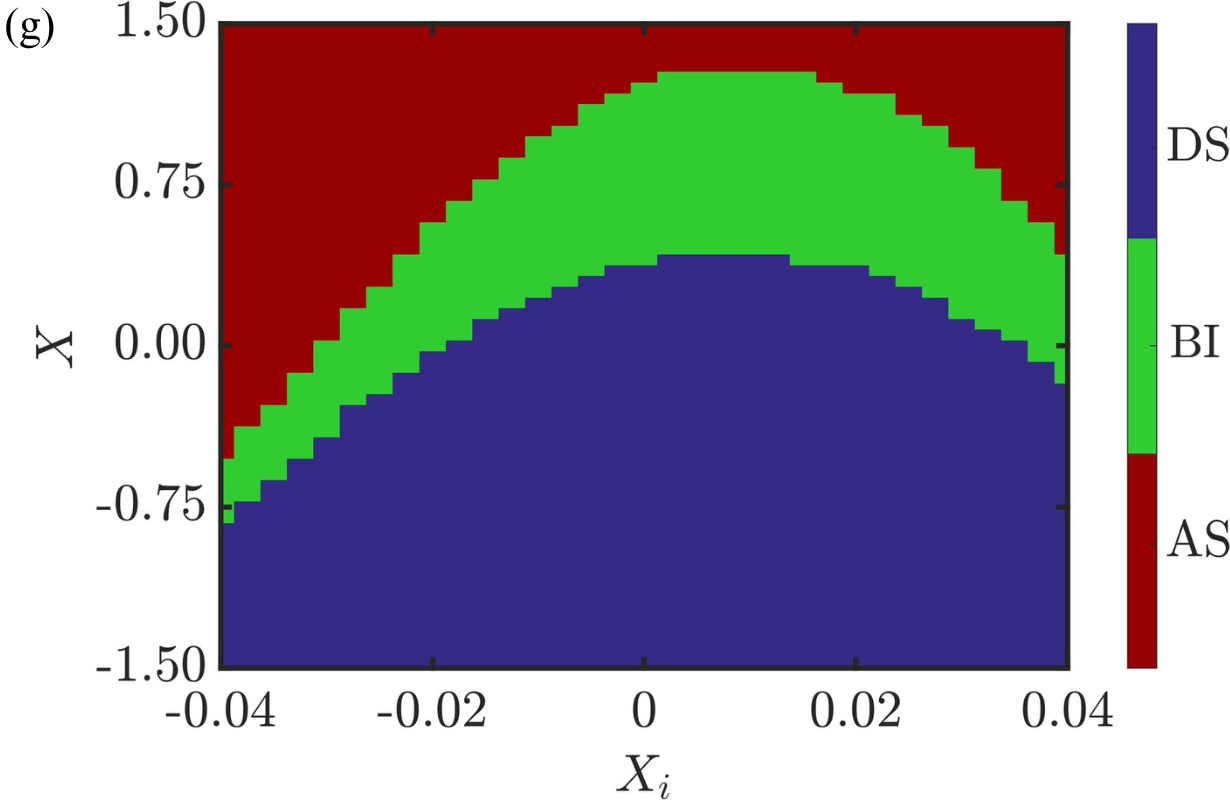}    
\end{minipage}
\caption{
\label{fig:tau_Xi_changeX}
Inhibitory neuronal heterogeneity modulates phase relations between cortical-like areas.
(a)-(f) The mean time delay $\tau$ between neuronal populations as a function of the parameter controlling the inhibitory neuronal heterogeneity $X_i$. The DS-AS transition can occur via a bistable regime for $g_I=4.00$~nS and $g_E=0.50$~nS for different values of $X$.
(g) Two-dimensional projections of the phase diagram of our model when varying the excitatory and inhibitory neuronal heterogeneity: DS (blue), AS (red), and bistability (BI, green). For this set of parameters, the transition from DS to AS only occurs via bistability. Each curve in (a)-(f) represents a horizontal line for a fixed $X$ in (g).
}
\end{figure}%

\begin{figure}[!ht]%
\begin{minipage}{8cm}
\includegraphics[width=0.96\columnwidth,clip]{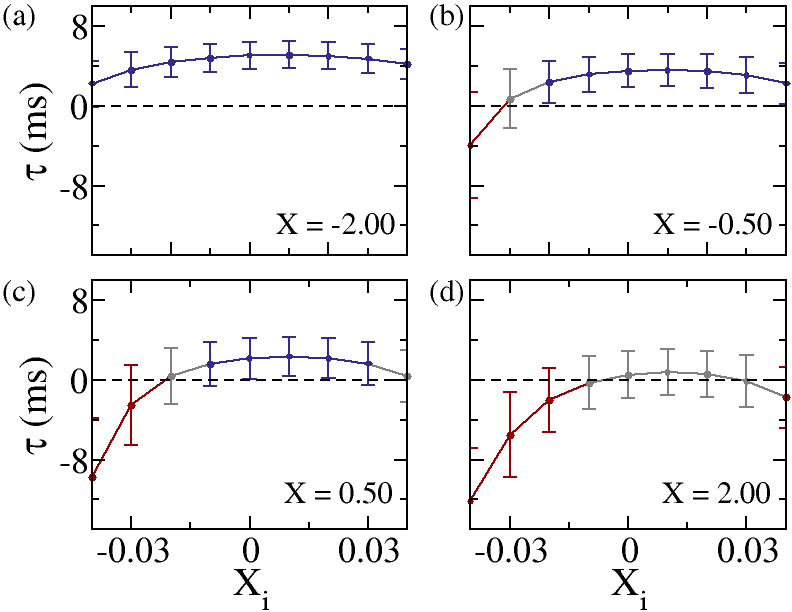}
\end{minipage}
\begin{minipage}{8cm}
\includegraphics[width=0.96\columnwidth,clip]{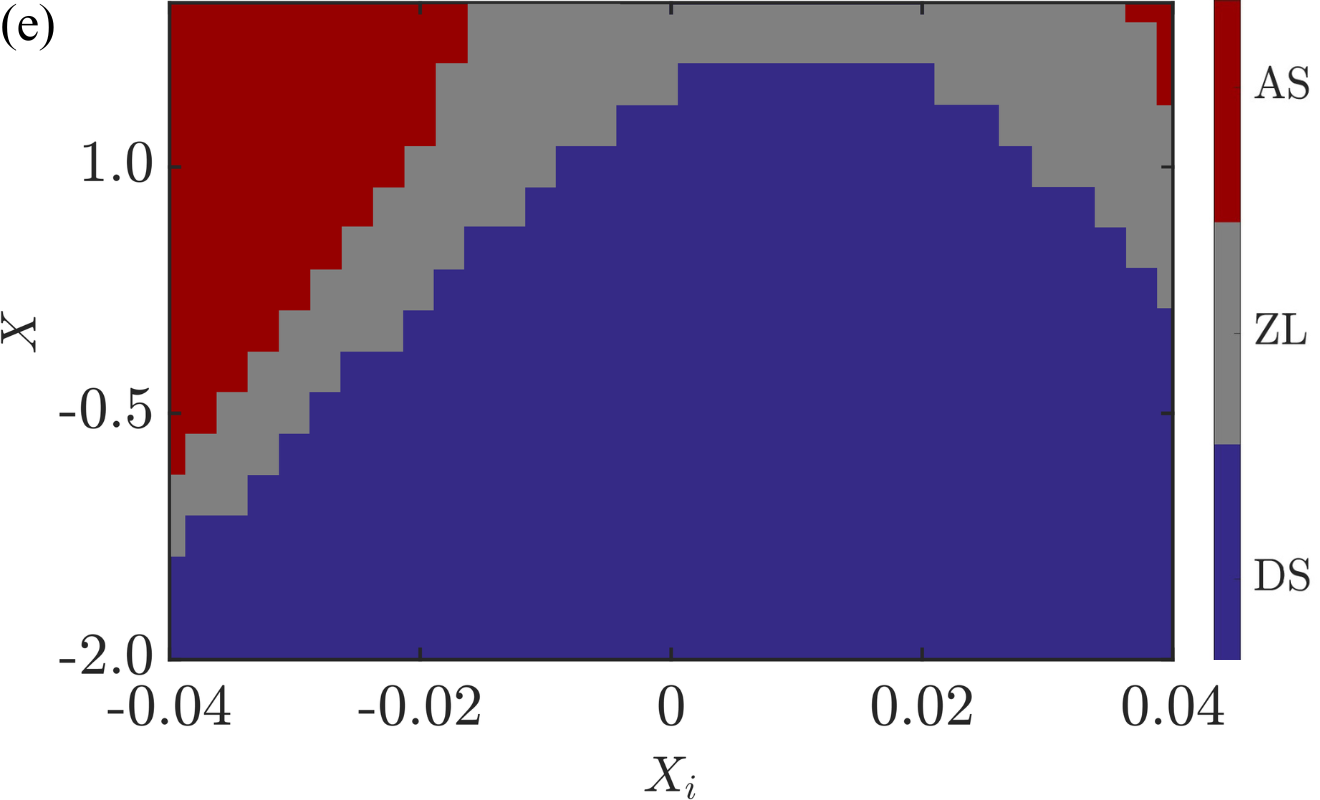}
\end{minipage}
\caption{\label{fig:tau_Xi_changeX_gi6} (a)-(d) The man time delay $\tau$ as a function of $X_i$ for $g_I = 6.00$~nS. The DS-AS transition occurs via a zero-lag regime between.
(e) Two-dimensional projections of the phase diagram of our model when varying the excitatory and inhibitory neurons: DS (blue), AS (red), and zero-lag (ZL, grey). The transition from DS to AS only occurs via ZL for this set of parameters. Each curve in (a)-(d) represents a horizontal line for a fixed $X$ in (e). }
\end{figure}%

\begin{figure}[!t]%
\begin{minipage}{8cm}
\includegraphics[width=1.0\columnwidth,clip]{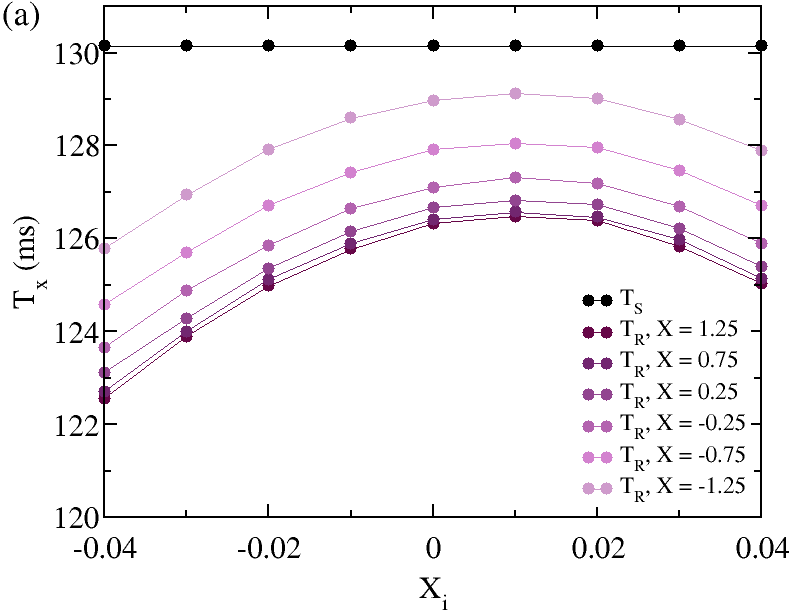}
\end{minipage}
\begin{minipage}{8cm}
\includegraphics[width=1.0\columnwidth,clip]{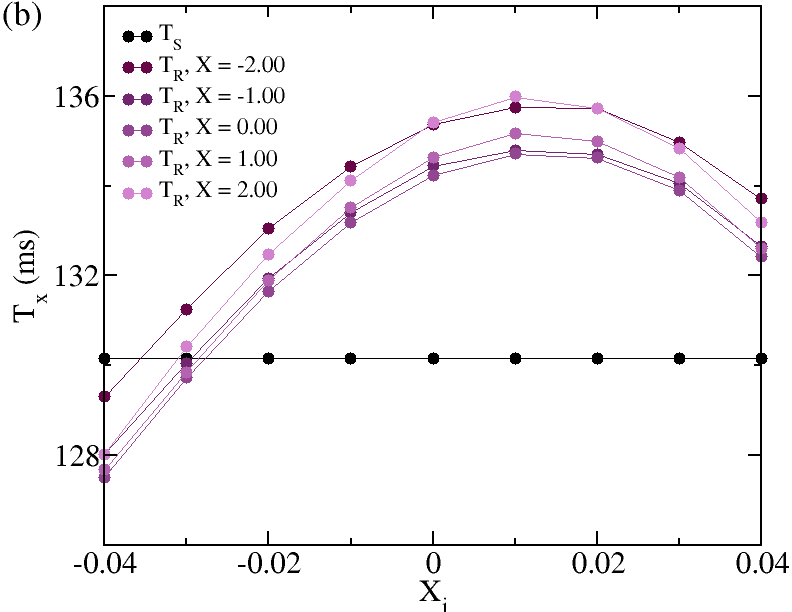}   
\end{minipage}
\caption{\label{fig:T_Xi_changeX}
Exploring the uncoupled situation ($g_E=0.0$~nS) and comparing the period of the sender with the period of the free-running receiver:  $T_x$ as a function of the parameter controlling the excitatory heterogeneity $X_i$.  
 (a) For $g_I=4.00$~nS,
 all the parameters are comparable to the ones in Fig.~\ref{fig:tau_Xi_changeX}, except the excitatory synaptic coupling $g_E$. We suggest that $T_R<T_S$ is an indication of the mechanism underlying the bistability.
 (b) For $g_I=6.00$~nS, 
 all the parameters are comparable to the ones in Fig.~\ref{fig:tau_Xi_changeX_gi6} except the excitatory synaptic coupling $g_E$.
 We suggest that the zero-lag transition is related to the region in which the colored lines cross the black one: $T_R=T_S$.
}
\end{figure}%

\begin{figure}[!t]%
\begin{minipage}{8cm}
\includegraphics[width=1.0\columnwidth,clip]{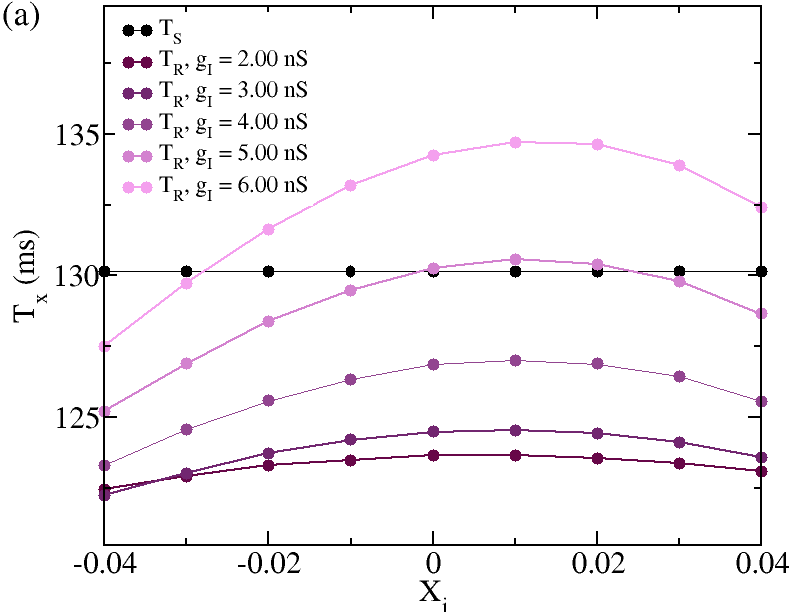}
\end{minipage}
\begin{minipage}{8cm}
\includegraphics[width=1.0\columnwidth,clip]{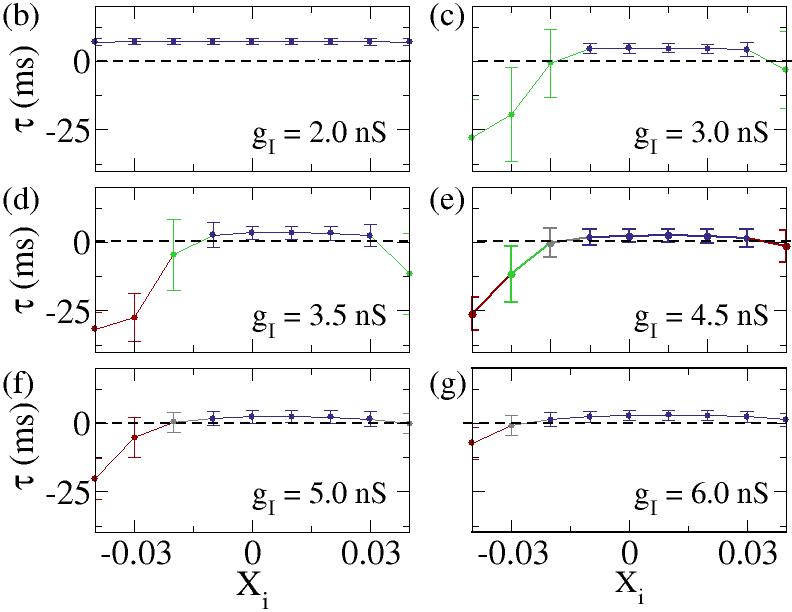}
\end{minipage}
\caption{\label{fig:tau_Xi_changegi}
Comparing the free-running period of the receiver with the period of the sender and their relation to the DS-AS transition via zero-lag or bistability. All plots show $X=0.0$~nS and different values of the inhibitory conductance ($g_I$). 
(a) The period of the population $T_x$ as a function of the parameter controlling the excitatory heterogeneity $X_i$, for the uncoupled situation ($g_E=0.0$~nS). 
(b)-(g) Mean time delay $\tau$ \textit{versus} $X_i$.
}
\end{figure}%

\subsection{Inhibitory neuronal heterogeneity modulates phase relations between populations}
  
The different patterns of phase relations between only-FS and only-LTS (specifically comparing Figs.~\ref{fig:FS_LTS_map_gi_X}(a) and (b)) suggest that networks with different proportions of these two types of neurons would also exhibit different phase relations. We investigate how changes in the distribution of different types of inhibitory neurons affect the coherent oscillations of the networks. In fact, the parameter $X_i$ controlling the proportion of FS and LTS neurons can modulate transitions from DS to AS. In what follows, we investigate the effect of incremental changes in the inhibitory neuronal heterogeneity $X_i$ (see neuronal distribution in Fig.~\ref{fig:neuronprobability}). The systems exhibit a rich repertoire of phase diversity depending on the combination of these three parameters: inhibitory synaptic conductance $g_i$, excitatory neuronal heterogeneity $X$, and inhibitory neuronal heterogeneity $X_i$. 

\subsubsection*{The DS-AS transition via bistability}

For a fixed value of  $g_i=4.0$~nS, we can investigate the dependence of the phase relation with both $X$ and $X_i$.
Figs.~\ref{fig:tau_Xi_changeX}(a)-(f) show the mean time delay $\tau$ as a function of $X_i$, for different values $X$.
Fig.~\ref{fig:tau_Xi_changeX}(g) displays a two-dimensional projection of the parameter space of our model ($X$,$X_i$). 
Each $\tau$ \textit{versus} $X_i$ curve in Figs.~\ref{fig:tau_Xi_changeX}(a)-(f) corresponds to a horizontal line in the colored map ~\ref{fig:tau_Xi_changeX}(g). For this specific value of inhibitory synaptic conductance $g_I=4.0$~nS 
the AS-DS transition occurs only via bistability as we increase $X_i$.
The system can also present a second transition from DS to AS, which is also mediated by a bistability in phase.

The inhibitory heterogeneity enlarges the region of bistability. For example, 
we can compare the width of the green region in a vertical line of $X_i= -0.04$ and $X_i= 0.1$ in Fig. ~\ref{fig:tau_Xi_changeX}(g). For
$X_i\approx 0.1$, which indicates more neurons in between FS and LTS, the system can present a bistable regime for many values of $X$, allowing different combinations of excitatory neurons with spiking patterns similar to IB.

\subsubsection*{The DS-AS transition via zero-lag}

For a larger value of inhibitory synaptic conductance $g_I=6.0$~nS, 
the AS-DS transitions occur only via zero-lag synchronization as we change $X_i$.
Fig.~\ref{fig:tau_Xi_changeX_gi6}(a)-(d) shows the mean time delay $\tau$ as a function of $X_i$, for different values $X$.
Fig.~\ref{fig:tau_Xi_changeX_gi6}(e) displays a two-dimensional projection of the parameter space ($X$,$X_i$). 
The system can also exhibit a second transition from DS to AS, which is also mediated by zero-lag. Each $\tau$ \textit{versus} $X_i$ curve in Figs.~\ref{fig:tau_Xi_changeX_gi6}(a)-(d) corresponds to a horizontal line in the colored map ~\ref{fig:tau_Xi_changeX_gi6}(e). 
One can also note that the presence of the bistability ensures that the regions of AS close to the AS-DS transition can present large anticipation time $\tau\approx-30$~ms, which does not occur when the transition is via zero-lag regime. This can be verified comparing Fig. ~\ref{fig:tau_Xi_changeX}(c) ~\ref{fig:tau_Xi_changeX_gi6}(c), for example.

\subsection{The mechanisms underlying the DS-AS transition: insights from the uncoupled situation}

The anticipated synchronization (AS) phenomenon can occur when the receiver's free-running dynamics are faster than the sender's~\cite {DallaPorta19,Hayashi16}. However, a complete mechanism explaining the transition from delayed synchronization (DS) to AS, whether through zero-lag synchronization (ZL) or bistability, remains to be discovered. To investigate this, we explore the system's uncoupled state ($g_E=0$). We calculate the mean period of each population $T_S$ and $T_R$ as we modify the parameter controlling the inhibitory heterogeneity $X_i$ only in the free-running receiver population.

For $g_I=4.0$~nS and different values of excitatory heterogeneity (ranging from $-1.25<X<1.25$) the receiver is always faster than the sender $T_R<T_S$ for all $-0.04<X_i<0.04$. The dependence of the periods with $X_i$ is shown in Fig.~\ref{fig:T_Xi_changeX}(a). $T_R$ behaves like a smooth parabolic function of $X_i$.

Except for the coupling, which is the excitatory synaptic conductance between the two populations $g_E$, all other parameters are the same in Figs. \ref{fig:T_Xi_changeX}(a) and 
~\ref{fig:tau_Xi_changeX}(a). Comparing both situations and figures, we can see that values of $X_i$ that maximize $T_R$ are related to larger values of time delays $\tau$ 

Regarding only the periods, it is not possible to predict if the system will be in DS or AS when coupled. However, if there is a transition, two things are consistent: (i) AS is in the regions where the difference $T_S  - T_R$ is larger (the transition occurs in the directions in which the receiver gets even faster); (ii) If $T_R$ is already smaller than $T_S$ in the DS regime, the lines do not cross, and the transition is via a bistable regime. 
We suggest that the transition via bistability is facilitated 
if $T_R<T_S$ already for the DS regime. This means that even in the DS regime, the receiver is faster than the sender, and there is no region of $T_R = T_S$. 

On the other hand, for a larger value of inhibitory synaptic conductances $g_I=6.0$~nS and different values of excitatory heterogeneity (ranging from $-2.0<X<2.0$), the receiver can be faster or slower than the sender depending on the inhibitory heterogeneity. 
The periods $T_R$ and $T_S$ as a function of $X_i$ for $g_I=6.0$~nS are shown in Fig.\ref{fig:T_Xi_changeX} (a). It is worth mentioning that for $X_i\approx -0.3$, the periods of both populations are similar ($T_R=T_S$ and the colored lines cross the black one).

In order to compare the coupled and uncoupled cases, see that all parameters in Figs.\ref{fig:T_Xi_changeX}(b) and 
~\ref{fig:tau_Xi_changeX_gi6}(a) are the same, except the coupling $g_E$.
When the system presents a DS in which the free-running receiver is slower than the sender ($T_R>T_S$), if we change parameters in direction to make the free-running faster, the transition from DS to AS can occur via zero lag. 
In this case, the transition via ZL occurs when the receiver's free-running period approaches the sender's, when lines cross each other $T_R=T_S$. It means that $\tau\approx0$ in the coupled situation for the values of $X_i$ in which $T_R\approx T_S$.

 We can study intermediate values of inhibitory conductance by fixing $X=0.0$ and comparing the coupled and uncoupled situations. 
 Fig.~\ref{fig:tau_Xi_changegi}(a) shows the period of the free-running situation as a function of $X_i$ for different values of $g_i$ and $g_E=0.0$~nS.
  In this case, we can see that depending on $g_I$, the $T_R$ colored curves can cross or not the $T_S$ black line. In fact, depending on the inhibitory synaptic conductances in the coupled situation, the transition from DS to AS can occur via zero-lag of bistability. Fig.~\ref{fig:tau_Xi_changegi}(b)-(g) show the time delay $\tau$ as a function of $X_i$ for the same parameters as in Fig.~\ref{fig:tau_Xi_changegi}(a) except that $g_E=0.5$~nS. Therefore, we suggest that the mechanism underlying the DS-AS transition depends on the relation between the free-running periods of the two populations.

\section{Concluding remarks}
\label{conclusions}

We demonstrated that inhibitory neuronal variability can drive different phase relationships between two spiking neuron networks. Specifically, we showed that both homogeneous and heterogeneous inhibitory networks can exhibit a diversity of phase relationships. By adjusting the proportion of fast-spiking (FS) neurons and low-threshold spiking (LTS) neurons, the system can exhibit phase-locking regimes with positive (DS), negative (AS), and zero-lag (ZL) phase differences, as well as a bistable regime (BI) between DS and AS. Moreover, we show that the DS-AS transition can occur via zero-lag synchronization or phase bistability and that the time delay is a function of the inhibitory properties. Our results also suggest that the mechanism underlying the DS-AS transition depends on the internal dynamics of the receiver population.

Therefore, our findings are significant in two complementary directions. On the one hand, our study contributes to exploring the biological mechanisms underlying the phenomena of anticipated synchronization, bistability, and phase diversity. On the other hand, our results shed light on the possible functional significance of inhibitory neuronal heterogeneity in the nervous system.

In the first direction, the DS-AS transition may explain frequently observed short-latency responses in visual systems~\cite{Orban85,Nowak95,Kerzel03,Jancke04,Puccini07,Martinez14}, olfactory circuits~\cite{Rospars14}, songbird brains~\cite{Dima18}, and human perception~\cite{Stepp10,Stepp17}. At the same time, the bistable regime can be related to bistable perception during ambiguous stimuli~\cite{Kosem16,machado2020phase}. However, previous studies on AS~\cite{Matias14,DallaPorta19} and phase bistability~\cite{machado2020phase,brito2021neuronal} did not explore the influence of inhibitory variability on network dynamics or the different mechanisms underlying the AS-DS transition via zero-lag synchronization or bistability.

In the second direction, while neuronal variability is a well-documented phenomenon across the brain, its role in information processing remains debated. Recent studies have increasingly focused on the functional roles of intrinsic neuronal heterogeneity in network dynamics~\cite{di2021optimal,Padmanabhan10,Savard11,Angelo12,Tripathy13}. The effects of heterogeneity on synchronization properties in single neuronal networks have been analyzed using various models~\cite{Golomb93,Perez10,Mejias12,Mejias14,rossi2020}. Here, we contribute to the field by studying how neuronal heterogeneity could affect phase relations between cortical areas and, consequently, communication between these areas.
 
Furthermore, unlike earlier studies on AS~\cite{Voss00,Ciszak03}, where anticipation time was predetermined by the dynamic equations or tied to excitation-inhibition interactions, our findings indicate that it can naturally emerge from heterogeneity. This opens new possibilities for studying how neuronal variability modulates phase relation diversity between cortical areas~\cite{Maris13,Maris16}. Finally, incorporating the effects of homeostasis, plasticity, and excitation-inhibition balance in our model are natural next steps we are currently pursuing.

\begin{acknowledgments}
The authors thank 
CNPq (grants 402359/2022-4, 314092/2021-8), FAPEAL (grant APQ2022021000015), UFAL, CAPES and L’ORÉAL-UNESCO-ABC For Women In Science (Para Mulheres na Ciência) for financial support.

\end{acknowledgments}
%

\bibliography{matias}

\end{document}